\documentclass[a4paper,11pt]{article}
\usepackage[utf8]{inputenc}
\usepackage[top=0.95in, bottom=0.75in, left=1.28in, right=0.8in]{geometry}
\usepackage{mathtools,hyperref}
\usepackage{tikz}
\usepackage{graphics}

\title{\large \textbf {Entanglement of Cavity Light Produced by a Superposed Two-Mode Coherent and Subharmonic Light}}
\author{\large Menwuyelet Melaku and Deribe Hirpo\\Department of Physics, Addis Ababa University\\P. O. Box 1176, Addis Ababa, Ethiopia\\
E-mail, {menwuyelet.melaku@aau.edu.et}}
\begin{document}

\maketitle

\begin{abstract}
With the aid of the solutions of the Quantum Langevin equations, we have calculated the Q-functions for the two-mode coherent and subharmonic cavity light beams. We have then determined the Q-function for the superposed two-mode coherent and subharmonic cavity light beams. With the help of the pertinent Q-functions for cavity light beams, we determined the photon statistics, quadrature statistics, and entanglement. we have obtained that the superposed two-mode coherent and subharmonic cavity light beams are in a squeezed state and the squeezing occurs in the plus quadrature. Besides, our analysis shows that at steady state and at threshold, the superposed two-mode coherent and subharmonic light beams have a maximum squeezing of $25\%$ below the two-mode vacuum-state level. We have also clearly shown that the  superposed two-mode light beam is entangled at steady-state and threshold and the entanglement turned out to be observed in the highly correlated squeezed photons with $75\%$ degree of entanglement.
\end{abstract}
\providecommand{\keywords}[1]{\textbf{{keywords:}} #1}
\keywords{Entanglement, Quadrature and photon statistics, Superposition of cavity modes}
\section{Introduction}
Quantum optics deals mainly with the quantum properties of the light generated by various optical systems such as lasers and with the effect of light on the dynamics of atoms. The quantum properties of light are largely determined by the states of the light mode and the well known quantum states of light are the number state, the chaotic state, the coherent state and the squeezed state [1]. Of all states of light, the coherent states are the most important and arise frequently in quantum optics. Not only can they be an accurate representation of the light produced by a stabilished laser operating well above threshold, but also many of the techniques for studying the properties of the light rely on the properties of the coherent states [2].

one of the most fundamentally interesting and intriguing phenomena associated with the composite quantum system is entanglement. In recent years, the topic of continuous-variable entanglement has received a significant amount of attention as it plays an important role in all branches of quantum information processing . The efficiency of quantum information schemes highly depends on the degree of entanglement [3].
\vspace {0.2cm}

Quantum entanglement is a physical phenomenon that occurs when pairs or groups of particles cannot be described independently instead, a quantum state may be given for the system as a whole. Measurements of physical properties such as position, momentum, spin polarization, etc performed on entangled particles are found to be appropriately correlated [3,6]. 
\vspace {0.2cm}

A pair of particles is taken to be entangled in quantum theory, if its states cannot be expressed as a product of the states of its individual constituents. The preparation and manipulation of these entangled states that have non-classical and non-local properties lead to better understanding of the basic quantum principles. In other words, it is a well-known fact that a quantum system is said to be entangled, if it is not separable. That is, if the density operator for the combined state cannot be described as a combination of the product density operators of the constituents [3,6,7].
\vspace {0.2cm}

A two-mode subharmonic generator at and above threshold has been theoretically predicted to be a source of light in an entangled state [8,9]. Recently, the experimental realization of the entanglement in two-mode subharmonic generator has been demonstared by Zhang et al [10]. They have found that a non-degenerate three-level laser can generate light in an entangled state employing the entanglement criteria for Bipartite continuous-variable state [11].
\vspace {0.2cm}

A detailed analysis of the quadrature squeezing and photon statistics of the light produced by a subharmonic generation has
been made by a number of authors. It has been shown theoretically and subsequently confirmed experimentally that subharmonic generation produces a light that has a maximum of $50\%$ squeezing below the coherent state level [11,12,13]. On the other hand, Xiong et al. have recently proposed a scheme for an entanglement based on a non-degenerate three-level laser when the three level atoms are injected at the lower level and the top and bottom levels are coupled by a strong coherent light. They have found that a non- degenerate three-level laser can generate light in entangled state employing the entanglement criteria for bipartite continuous-variable states[6,11].
\vspace {0.2cm}

Even though Einstein, along with his colleagues Podolsky and Rosen, was first to recognize the criterion for analyzing entanglement condition for a two-mode light beams [14] , a significant number of works have not been devoted on a pair of superposed two-mode light beams.
\vspace {0.1cm}

Hence, in this article we seek to analyze the statistical, quadrature and entanglement properties of a superposed two-mode coherent and subharmonic cavity light beams. To this end, With the aid of Quantum Langevin equations and operator dynamics, we calculate the Q-function for the two-mode coherent and subharmonic cavity light. We then determine the Q-function for the superposed two-mode coherent and subharmonic cavity light beams. With the help of the pertinent Q-functions, we determine the cavity photon quadrature statistics and entanglement for the superposed two-mode coherent and subharmonic cavity light beams.

\section{Two-Mode Coherent Light}
In this section we consider a cavity mode driven by a two-mode coherent light and coupled to a two-mode vacuum reservoir via a single port mirror. The interaction between a two-mode cavity light and a two-mode driving coherent light can be described by a Hamiltonian
\begin{equation}\label{1}
\hat H=i\varepsilon(\hat a^\dagger-\hat a+\hat b^\dagger-\hat b),
\end{equation}
where $\hat a$($\hat b$) is the annihilation operator for the cavity modes $a(b)$ and $\varepsilon$ is proportional to the amplitude of the driving modes. Applying Eq.~\eqref{1}, the master equation for a cavity mode driven by a two-mode coherent light and coupled to a two-mode vacuum reservor, can be written as [1]
\begin {eqnarray}\label{2}
\frac{d\hat \rho}{dt}&=&-\varepsilon(\hat a\hat \rho-\hat a^\dagger\hat \rho-\hat\rho\hat a+\hat \rho\hat a^\dagger+\hat b\hat \rho-\hat b^\dagger\hat \rho-\hat\rho\hat b+\hat \rho\hat b^\dagger)\nonumber\\
&+&\frac{\kappa}{2}(2\hat a\hat\rho\hat a^\dagger-\hat a^\dagger\hat a\hat\rho-\hat\rho\hat a^\dagger\hat a)+\frac{\kappa}{2}(2\hat b\hat\rho\hat b^\dagger-\hat b^\dagger\hat b\hat\rho-\hat\rho\hat b^\dagger\hat b).
\end {eqnarray}
Employing a relation,
\begin{equation}\label{3}
\frac{d}{dt}\langle \hat A\rangle=Tr(\frac{d\hat\rho}{dt}\hat A),
\end{equation}
along with Eq.~\eqref{2}, we have
\begin {eqnarray}\label{4}
\frac{d}{dt}\langle \hat a(t)\rangle&=&-\varepsilon Tr(\hat a\hat \rho\hat a-\hat a^\dagger\hat \rho\hat a-\hat\rho\hat a^2+\hat \rho\hat a^\dagger\hat a+\hat b\hat \rho\hat a-\hat b^\dagger\hat \rho\hat a-\hat\rho\hat b\hat a+\hat \rho\hat b^\dagger\hat a)\nonumber\\
&+&\frac{\kappa}{2}Tr(2\hat a\hat\rho\hat a^\dagger\hat a-\hat a^\dagger\hat a\hat\rho\hat a-\hat\rho\hat a^\dagger\hat a^2)+\frac{\kappa}{2}Tr(2\hat b\hat\rho\hat b^\dagger\hat a-\hat b^\dagger\hat b\hat\rho\hat a-\hat\rho\hat b^\dagger\hat b\hat a).
\end {eqnarray}
Applying a cyclic property of trace operator along with the commutation relation
\begin{equation}\label{5}
[\hat a,\hat a^\dagger]=[\hat b,\hat b^\dagger]=1,
\end{equation}
and 
\begin{equation}\label{6}
[\hat a,\hat b]=[\hat a,\hat b^\dagger]=[\hat a^\dagger,\hat b]=[\hat a^\dagger,\hat b^\dagger]=0,
\end{equation}
we readily get
\begin{equation}\label{7}
\frac{d}{dt}\langle \hat a(t)\rangle=-\frac{\kappa}{2}\langle \hat a(t)\rangle+\varepsilon.
\end{equation}
Similarly,
\begin{equation}\label{8}
\frac{d}{dt}\langle \hat b(t)\rangle=-\frac{\kappa}{2}\langle \hat b(t)\rangle+\varepsilon,
\end{equation}
\begin{equation}\label{9}
\frac{d}{dt}\langle \hat a(t)\hat b(t)\rangle=-\kappa\langle \hat a(t)\hat b(t)\rangle+\varepsilon(\langle \hat a(t)\rangle+\langle \hat b(t)\rangle),
\end{equation}
\begin{equation}\label{10}
\frac{d}{dt}\langle \hat a^2(t)\rangle=-\kappa\langle \hat a^2(t)\rangle+2\varepsilon\langle \hat a(t)\rangle,
\end{equation}
\begin{equation}\label{11}
\frac{d}{dt}\langle \hat a^\dagger(t)\hat a(t)\rangle=-\kappa\langle \hat a^\dagger(t)\hat a(t)\rangle+\varepsilon(\langle \hat a^\dagger(t)\rangle+\langle \hat a(t)\rangle),
\end{equation}
\begin{equation}\label{12}
\frac{d}{dt}\langle \hat a^\dagger(t)\hat b(t)\rangle=-\kappa\langle \hat a^\dagger(t)\hat b(t)\rangle+\varepsilon(\langle \hat a^\dagger(t)\rangle+\langle \hat b(t)\rangle).
\end{equation}
Using Eqs.~\eqref{7} and ~\eqref{8} we can write
\begin{equation}\label{13}
\frac{d}{dt} \hat a(t)=-\frac{\kappa}{2} \hat a(t)+\varepsilon+\hat f_a(t),
\end{equation}
\begin{equation}\label{14}
\frac{d}{dt}\hat b(t)=-\frac{\kappa}{2}\hat b(t)+\varepsilon+\hat f_b(t).
\end{equation}
Taking the expectation value of the above two equations and comparing with Eqs.~\eqref{7} and ~\eqref{8} yields
\begin{equation}\label{15}
\langle\hat f_a(t)\rangle=\langle\hat f_b(t)\rangle=0.
\end{equation}
The formal solution of Eqs.~\eqref{13} and ~\eqref{14} can be expressible as
\begin{equation}\label{16}
\hat a(t)=\hat a(0)e^{-\kappa t/2}+\int^t_0e^{-\kappa(t-t')/2}[\varepsilon+\hat f_a(t')]dt',
\end{equation}
\begin{equation}\label{17}
\hat b(t)=\hat b(0)e^{-\kappa t/2}+\int^t_0e^{-\kappa(t-t')/2}[\varepsilon+\hat f_b(t')]dt'.
\end{equation}
Which can also be rewritten in the form
\begin{equation}\label{18}
\hat a(t)=p(t)\hat a(0)+q(t)+f_+(t)+f_-(t),
\end{equation}
\begin{equation}\label{19}
\hat b(t)=p(t)\hat b(0)+q(t)+f^\dagger_+(t)-f^\dagger_-(t),
\end{equation}
where,
\begin{equation}\label{20}
p(t)=e^{-\kappa t/2},
\end{equation}
\begin{equation}\label{21}
q(t)=\frac{2\varepsilon}{\kappa}\bigg\lgroup1-e^{-\kappa t/2}\bigg\rgroup,
\end{equation}
and
\begin{equation}\label{22}
f_\pm(t)=\frac{1}{2}\int^t_0e^{-\kappa(t-t')/2}[\hat f_a(t')\pm\hat f^\dagger_b(t')]dt'.
\end{equation}
One can write Eqs.~\eqref{18} and ~\eqref{19} as
\begin{equation}\label{23}
\hat a(t)=\hat a'(t)+q(t),
\end{equation}
\begin{equation}\label{24}
\hat b(t)=\hat b'(t)+q(t),
\end{equation}
with
\begin{equation}\label{25}
\hat a'(t)=p(t)\hat a(0)+\int^t_0e^{-\kappa(t-t')/2}\hat f_a(t')dt',
\end{equation}
\begin{equation}\label{26}
\hat b'(t)=p(t)\hat b(0)+\int^t_0e^{-\kappa(t-t')/2}\hat f_b(t')dt'.
\end{equation}
Now we need to show that $\hat a'(t)$ and $\hat b'(t)$ are Gaussian variables. To this end taking expectation value of Eqs.~\eqref{25} and ~\eqref{26} and differentiating with respect to t yields
\begin{equation}\label{27}
\frac{d}{dt}\langle \hat a'(t)\rangle=-\frac{\kappa}{2}\langle \hat a'(t)\rangle,
\end{equation}
\begin{equation}\label{28}
\frac{d}{dt}\langle \hat b'(t)\rangle=-\frac{\kappa}{2}\langle \hat b'(t)\rangle.
\end{equation}
which are linear equations, we then see that $\hat a'(t)$ and $\hat b'(t)$ are Gaussian variables. In addition, using Eqs.~\eqref{25} and ~\eqref{26} and assuming that the cavity mode is initially in a two-mode vacuum state, we observe that
\begin{equation}\label{29}
\langle \hat a'(t)\rangle=\langle \hat b'(t)\rangle=0.
\end{equation}
Hence $\hat a'(t)$ and $\hat b'(t)$ are Gaussian variables with vanishing means.
We now proceed to find Q-function for a two-mode coherent cavity light. In general Q-function for a two-mode light is expressible as 
\begin{equation}\label{30}
Q(\alpha,\beta, t)=\frac{1}{\pi^4}\int d^2zd^2\eta \phi_a(z,\eta, t)exp(z^*\alpha+\eta^*\beta-z\alpha^*-\eta\beta^*),
\end{equation}
where, $\phi_a(z,\eta,t)$ is a characterstic function described by 
\begin{equation}\label{31}
\phi_a(z,\eta,t)=Tr(\hat \rho(0)e^{-z^*\hat a(t)}e^{z\hat a^\dagger(t)}e^{-\eta^*\hat b(t)}e^{\eta\hat b^\dagger(t)}).
\end{equation}
Employing an identity,
\begin{equation}\label{32}
e^{\hat A}e^{\hat B}=e^{\hat A+\hat B+\frac{1}{2}[\hat A, \hat B]},
\end{equation}
along with Eqs.~\eqref{23} and ~\eqref{24}, we obtain
\begin{eqnarray}\label{33}
\phi_a(z,\eta,t)&=&exp[-\frac{1}{2} z^*z-\frac{1}{2}\eta^*\eta+(z-z^{*}+\eta-\eta^{*})q(t)]\langle exp[z\hat a^{'\dagger}(t)-z^*\hat a^{'}(t)\nonumber\\
&+&\eta \hat b^{'\dagger}(t)-\eta^{*}\hat b^{'}(t)]\rangle.
\end{eqnarray}
We know that $\hat a'(t)$ and $\hat b'(t)$ are Gaussian operators with zero mean. Hence, using a relation 
\begin{equation}\label{34}
\langle exp(\hat a(t)+\hat b(t))\rangle=exp[\frac{1}{2}\langle (\hat a(t)+\hat b(t))^2\rangle],
\end{equation}
which holds true for Gaussian operators $\hat a(t)$ and $\hat b(t)$ with a vanishing mean, one can have
\begin{eqnarray}\label{35}
\phi_a(z,\eta,t)&=&exp[-\frac{1}{2} z^*z-\frac{1}{2}\eta^*\eta+(z-z^{*}+\eta-\eta^{*})q(t)] exp[-\frac{1}{2}z^*z\langle (\hat a^{'\dagger}(t)\hat a'(t)\nonumber\\
&+&\hat a'(t)\hat a^{'\dagger}(t))\rangle-\frac{1}{2}\eta^*\eta\langle (\hat b^{'\dagger}(t)\hat b'(t)+\hat b'(t)\hat b^{'\dagger}(t))\rangle+\frac{1}{2}z\eta\langle (\hat a^{'\dagger}(t)\hat b^{'\dagger}(t)\nonumber\\
&+&\hat b^{'\dagger}(t)\hat a^{'\dagger}(t))+\frac{1}{2}z^*\eta^*\langle (\hat a^{'}(t)\hat b'(t)+\hat b'(t)\hat a^{'}(t))\rangle-\frac{1}{2}z\eta^*\langle (\hat a^{'\dagger}(t)\hat b'(t)\nonumber\\
&+&\hat b'(t)\hat a^{'\dagger}(t))\rangle-\frac{1}{2}z^*\eta\langle (\hat a^{'}(t)\hat b^{'\dagger}(t)+\hat b^{'\dagger}(t)\hat a^{'}(t))\rangle+\frac{1}{2}z^2\langle\hat a^{'\dagger 2}(t)\rangle\nonumber\\
&+&\frac{1}{2}z^{*2}\langle\hat a^{'2}(t)\rangle+\frac{1}{2}\eta^2\langle\hat b^{'\dagger 2}(t)\rangle+\frac{1}{2}\eta^{*2}\langle\hat b^{'2}(t)\rangle].                                                                   
\end{eqnarray}
Now let's proceed to find the correlations of $\hat a'(t)$ and $\hat b'(t)$ which appeared in the above equation. To this end, using Eq.~\eqref{25} and its conjugate along with an assumption that the cavity mode is initially in a two-mode vacuum state and noting that a noise operator at later time doesn't affect cavity mode operator at earlier time, we may have
\begin {eqnarray}\label{36}
\langle\hat a'^\dagger(t)\hat a'(t)\rangle=\int^t_0e^{-\kappa(2t-t'-t'')/2}\langle\hat f^\dagger_a(t'')\hat f_a(t')\rangle dt'dt''.
\end {eqnarray}
Employing a relation,
\begin{equation}\label{37}
\frac{d}{dt}\langle \hat a^\dagger(t)\hat a(t)\rangle=\langle (\frac{d\hat a^\dagger(t)}{dt})\hat a(t)\rangle+\langle \hat a^\dagger(t)(\frac{d\hat a(t)}{dt})\rangle,
\end{equation}
along with Eq.~\eqref{13} and its conjugate we obtain
\begin{equation}\label{38}
\frac{d}{dt}\langle \hat a^\dagger(t)\hat a(t)\rangle=-\kappa\langle \hat a^\dagger(t)\hat a(t)\rangle+\varepsilon(\langle \hat a^\dagger(t)\rangle+\langle \hat a(t)\rangle)+\langle \hat f^\dagger_a(t)\hat a(t)\rangle+\langle \hat a^\dagger(t)\hat f_a(t)\rangle,
\end{equation}
comparing with Eq.~\eqref{11}, we observe that
\begin{equation}\label{39}
\langle \hat f^\dagger_a(t)\hat a(t)\rangle+\langle \hat a^\dagger(t)\hat f_a(t)\rangle=0.
\end{equation}
Multiplying Eq.~\eqref{16} with $\hat f^\dagger_a(t)$ from the left and conjugate of Eq.~\eqref{16} by $\hat f_a(t)$ from the right and taking expectation value of the resulting equation yields
\begin{equation}\label{40}
 \langle \hat f^\dagger_a(t)\hat a(t)\rangle=\langle \hat f^\dagger_a(t)\hat a(0)\rangle e^{-\kappa t/2}+\int^t_0 e^{-\kappa(t-t')/2}[\varepsilon\langle \hat f^\dagger_a(t)\rangle+\langle\hat f^\dagger_a(t)\hat f_a(t')\rangle]dt',
\end{equation}
\begin{equation}\label{41}
 \langle \hat a^\dagger(t)\hat f_a(t)\rangle=\langle \hat a^\dagger(0)\hat f_a(t)\rangle e^{-\kappa t/2}+\int^t_0 e^{-\kappa(t-t')/2}[\varepsilon\langle \hat f_a(t)\rangle+\langle\hat f^\dagger_a(t')\hat f_a(t)\rangle]dt',
\end{equation}
using the fact that noise operator has a vanishing mean and a noise operator at later time doesn't affect a cavity mode operator at earlier time 
\begin{equation}\label{42}
\langle \hat f^\dagger_a(t)\hat a(0)\rangle=\langle \hat f^\dagger_a(t)\rangle\langle\hat a(0)\rangle=0,
\end{equation}
\begin{equation}\label{43}
\langle \hat a^\dagger(0)\hat f_a(t)\rangle=\langle \hat a^\dagger(0)\rangle\langle\hat f_a(t)\rangle=0,
\end{equation}
Eqs.~\eqref{40} and ~\eqref{41} reduces to
\begin{equation}\label{44}
 \langle \hat f^\dagger_a(t)\hat a(t)\rangle=\int^t_0 e^{-\kappa(t-t')/2}\langle\hat f^\dagger_a(t)\hat f_a(t')\rangle dt',
\end{equation}
\begin{equation}\label{45}
 \langle \hat a^\dagger(t)\hat f_a(t)\rangle=\int^t_0 e^{-\kappa(t-t')/2}\langle\hat f^\dagger_a(t')\hat f_a(t)\rangle dt'.
\end{equation}
Adding Eqs.~\eqref{44} and ~\eqref{45}, inview of Eq.~\eqref{39}, and assuming 
\begin{equation}\label{46}
 \langle\hat f^\dagger_a(t)\hat f_a(t')\rangle =\langle\hat f^\dagger_a(t')\hat f_a(t)\rangle,
\end{equation}
we obtain 
\begin{equation}\label{47}
 \int^t_0 e^{-\kappa(t-t')/2}\langle\hat f^\dagger_a(t')\hat f_a(t)\rangle dt'=0.
\end{equation}
Using a relation,
\begin{equation}\label{48}
 \int^t_0 e^{-a(t-t')}\langle F(t) G(t')\rangle dt'=b,
\end{equation}
we assert that
\begin{equation}\label{49}
\langle F(t) G(t')\rangle=2b\delta(t-t').
\end{equation}
Thus, we have
\begin{equation}\label{50}
 \langle\hat f^\dagger_a(t')\hat f_a(t'')\rangle=0,
\end{equation}
Substituting the above result in Eq.~\eqref{36} yields
\begin{equation}\label{51}
\langle \hat a'^\dagger(t)\hat a'(t)\rangle=0,
\end{equation}
Similarly,
\begin{equation}\label{52}
\langle \hat b'^\dagger(t)\hat b'(t)\rangle=0.
\end{equation}
Inview of Eq.~\eqref{25} along with its conjugate we also have
\begin {eqnarray}\label{53}
\langle\hat a'(t)\hat a'^\dagger(t)\rangle=p^2(t)+e^{-\kappa t}\int^t_0e^{\kappa(t'+t'')/2}\langle\hat f_a(t'')\hat f^\dagger_a(t')\rangle dt'dt''.
\end {eqnarray}
Making use of a relation,
\begin{equation}\label{54}
\frac{d}{dt}\langle \hat a(t)\hat a^\dagger(t)\rangle=\langle (\frac{d\hat a(t)}{dt})\hat a^\dagger(t)\rangle+\langle \hat a(t)(\frac{d\hat a^\dagger(t)}{dt})\rangle,
\end{equation}
together with Eq.~\eqref{13} and its conjugate we obtain
\begin{equation*}
\frac{d}{dt}\langle \hat a(t)\hat a^\dagger(t)\rangle=-\kappa\langle \hat a(t)\hat a^\dagger(t)\rangle+\varepsilon(\langle \hat a^\dagger(t)\rangle+\langle \hat a(t)\rangle)+\langle \hat f_a(t)\hat a^\dagger(t)\rangle+\langle \hat a(t)\hat f^\dagger_a(t)\rangle,
\end{equation*}
using a commutation relation described by Eq.~\eqref{5} the above equation can be rewritten as
\begin{equation}\label{55}
\frac{d}{dt}\langle \hat a^\dagger(t)\hat a(t)\rangle=-\kappa\langle \hat a^\dagger(t)\hat a(t)\rangle+\varepsilon(\langle \hat a^\dagger(t)\rangle+\langle \hat a(t)\rangle)+\langle \hat f_a(t)\hat a^\dagger(t)\rangle+\langle \hat a(t)\hat f^\dagger_a(t)\rangle-\kappa,
\end{equation}
comparing with Eq.~\eqref{11} we observe that
\begin{equation}\label{56}
\langle \hat f_a(t)\hat a^\dagger(t)\rangle+\langle \hat a(t)\hat f^\dagger_a(t)\rangle=\kappa.
\end{equation}
which gives
\begin{equation}\label{57}
 \langle\hat f_a(t')\hat f^\dagger_a(t'')\rangle=\kappa\delta(t'-t''),
\end{equation}
Replacing the above result in Eq.~\eqref{53} and making use of Eq.~\eqref{20} one can get
\begin{equation}\label{58}
\langle \hat a'(t)\hat a'^\dagger(t)\rangle=1,
\end{equation}
Similarly,
\begin{equation}\label{59}
\langle \hat b'(t)\hat b'^\dagger(t)\rangle=1.
\end{equation}
Following a similar procedure as above one can also estabilish for
\begin{equation}\label{60}
\langle\hat a^{'\dagger}(t)\hat b^{'\dagger}(t)\rangle=\langle\hat b^{'\dagger}(t)\hat a^{'\dagger}(t)\rangle=0,
\end{equation}
\begin{equation}\label{61}
\langle \hat a^{'}(t)\hat b'(t)\rangle=\langle\hat b'(t)\hat a^{'}(t)\rangle=0,
\end{equation}
\begin{equation}\label{62}
\langle\hat a^{'\dagger}(t)\hat b'(t)\rangle=\langle\hat b'(t)\hat a^{'\dagger}(t)\rangle=0,
\end{equation}
\begin{equation}\label{63}
\langle\hat a^{'}(t)\hat b^{'\dagger}(t)\rangle=\langle\hat b^{'\dagger}(t)\hat a^{'}(t)\rangle=0,
\end{equation}
\begin{equation}\label{64}
\langle\hat a^{'\dagger 2}(t)\rangle=\langle\hat a^{'2}(t)\rangle=0,
\end{equation}
\begin{equation}\label{65}
\langle\hat b^{'\dagger 2}(t)\rangle=\langle\hat b^{'2}(t)\rangle=0.
\end{equation}
Thus, using the above results the characterstic function described in Eq.~\eqref{35} reduces to
\begin{eqnarray}\label{66}
\phi_a(z,\eta,t)=exp[-z^*z-\eta^*\eta+(z-z^{*}+\eta-\eta^{*})q(t)].                                                                    
\end{eqnarray}
Introducing Eq.~\eqref{66} in ~\eqref{30} we have
\begin{eqnarray}\label{67}
Q(\alpha,\beta, t)&=&\frac{1}{\pi^2}\int \frac{d^2zd^2\eta}{\pi^2} exp[-z^*z-\eta^*\eta+(q-\alpha^*)z+(\alpha-q)z^{*}\nonumber\\
&+&(q-\beta^*)\eta+(\beta-q)\eta^{*}].
\end{eqnarray}
Employing a relation,
\begin{eqnarray}\label{68}
\int \frac{d^2z}{\pi}exp[-az^*z+bz+cz^*+Az^2+Bz^{*2}]\nonumber\\
=\bigg[\frac{1}{a^2-4AB}\bigg]^{1/2}exp \bigg\lgroup \frac{abc+Ac^2+Bb^2}{a^2-4AB}\bigg\rgroup, a>0,
\end{eqnarray}
and performing the integration, we readily obtain
\begin{equation}\label{69}
Q(\alpha,\beta, t)=\frac{1}{\pi^2}exp[-\alpha^*\alpha-\beta^*\beta+(\alpha+\beta+\alpha^*+\beta^*)q-2q^2].
\end{equation}
\section{Two-Mode Subharmonic Light}
In a two-mode subharmonic generator a pump-photon of frequency $\omega_c$ is down converted into highly correlated signal and idler photons with frequency $\omega_a$ and $\omega_b$ such that $\omega_c=\omega_a+\omega_b$ .
The process of two-mode Subharmonic generation is described by a Hamiltonian
\begin{equation}\label{70}
\hat H=i\lambda(\hat a\hat b\hat c^\dagger-\hat a^\dagger\hat b^\dagger\hat c),
\end{equation}
where, $\hat a(\hat b)$ is annihilation operator for the signal(idler) mode and $\hat c$ is annihilation operator for the pump mode, with $\lambda$ being the coupling constant.
\par
With the pump mode represented by a real and constant c-number $\mu$, the Hamiltonian can be rewritten as 
\begin{equation}\label{71}
\hat H=i\gamma(\hat a\hat b-\hat a^\dagger\hat b^\dagger),
\end{equation}
with, $\gamma=\lambda\mu$.
\par
With the help of a master equation for a cavity light interacting with a reservoir
\begin{equation}\label{72}
\frac{d\hat \rho}{dt}=-i[\hat H,\rho]+\frac{\kappa}{2}(\bar n+1)(2\hat a\rho \hat a^+-\hat a^+\hat a\hat \rho-\hat \rho\hat a^+\hat a)+\frac{\kappa}{2}\bar n(2\hat a^+\rho \hat a-\hat a\hat a^+\hat \rho-\hat \rho\hat a\hat a^+),
\end{equation}
along with a relation, given in Eq.~\eqref{3} and with a consideration that the cavity is coupled to a two-mode vacuum reservoir ($\bar n=0$) via a single-port mirror, the equation of evolution of the operators is evaluated as,
\begin {eqnarray}\label{73}
\frac{d}{dt}\langle \hat a\rangle&=&-i\langle[\hat a,\hat H]\rangle+\frac{\kappa}{2}Tr[(2\hat a\rho \hat a^{\dagger}-\hat a^\dagger\hat a\hat \rho-\hat \rho\hat a^\dagger\hat a)\hat a].
\end {eqnarray}
With the aid of the cyclic property of trace and the commutation relation described in Eqs.~\eqref{5} and ~\eqref{6}, we obtain
\begin{equation}\label{74}
\frac{d}{dt}\langle\hat a(t)\rangle=-\frac{1}{2}\kappa \langle\hat a(t)\rangle-\gamma\langle\hat b^\dagger(t)\rangle,
\end{equation}
using a similar procedure one can show that
\begin{equation}\label{75}
\frac{d}{dt}\langle\hat b(t)\rangle=-\frac{1}{2}\kappa \langle\hat b(t)\rangle-\gamma\langle\hat a^\dagger(t)\rangle,
\end{equation}
\begin {eqnarray}\label{76}
\frac{d}{dt}\langle \hat a^\dagger(t)\hat a(t)\rangle=-\kappa\langle \hat a^\dagger(t)\hat a(t)\rangle-\gamma(\langle\hat a(t)\hat b(t)\rangle+\langle \hat a^\dagger(t)\hat b^\dagger(t)\rangle),
\end {eqnarray}
\begin {eqnarray}\label{77}
\frac{d}{dt}\langle \hat b^\dagger(t)\hat b(t)\rangle=-\kappa\langle \hat b^\dagger(t)\hat b(t)\rangle-\gamma(\langle\hat a(t)\hat b(t)\rangle+\langle \hat a^\dagger(t)\hat b^\dagger(t)\rangle),
\end {eqnarray}
\begin {eqnarray}\label{78}
\frac{d}{dt}\langle \hat a(t)\hat b(t)\rangle=-\kappa\langle \hat a(t)\hat b(t)\rangle-\gamma(\langle\hat a^\dagger(t)\hat a(t)\rangle+\langle \hat b^\dagger(t)\hat b(t)\rangle)-\gamma,
\end {eqnarray}
\begin {eqnarray}\label{79}
\frac{d}{dt}\langle \hat a^\dagger(t)\hat b(t)\rangle=-\kappa\langle \hat a^\dagger(t)\hat b(t)\rangle-\gamma(\langle\hat a^{\dagger2}(t)\rangle+\langle \hat b^{\dagger2}(t)\rangle),
\end {eqnarray}
\begin {eqnarray}\label{80}
\frac{d}{dt}\langle \hat a^2(t)\rangle=-\kappa\langle \hat a^2(t)\rangle-2\gamma\langle\hat a(t)\hat b^{\dagger}(t)\rangle.
\end {eqnarray}
Introducing noise operators we can rewrite Eqs.~\eqref{74} and ~\eqref{75} as
\begin{equation}\label{81}
\frac{d}{dt}\hat a(t)=-\frac{1}{2}\kappa \hat a(t)-\gamma\hat b^\dagger(t)+\hat F_a(t),
\end{equation}
\begin{equation}\label{82}
\frac{d}{dt}\hat b(t)=-\frac{1}{2}\kappa \hat b(t)-\gamma\hat a^\dagger(t)+\hat F_b(t).
\end{equation}
We now proceed to find a solution for the above two- equations, to this end, let 
\begin{equation}\label{83}
\hat Z_\pm(t)=\hat a(t)\pm\hat b^\dagger(t),
\end{equation}
differentiating both sides of the above equation with respect to t, we have
\begin{equation}\label{84}
\frac{d}{dt}\hat Z_\pm(t)=\frac{d}{dt}\hat a(t)\pm\frac{d}{dt}\hat b^\dagger(t).
\end{equation}
Using Eq.~\eqref{81} and conjugate of Eq.~\eqref{82} into Eq.~\eqref{84} we see that
\begin{equation}\label{85}
\frac{d}{dt}\hat Z_\pm(t)=-\frac{1}{2}\lambda_\pm\hat Z_\pm(t)+\hat F_a(t)\pm\hat F^\dagger_b(t),
\end{equation}
where,
\begin{equation}\label{86}
\lambda_\pm=\kappa\pm2\gamma.
\end{equation}
The formal solution of Eq.~\eqref{85} is expressible as 
\begin{equation}\label{87}
\hat Z_\pm(t)=\hat Z_\pm(0)e^{-\lambda_\pm t/2}+\int^t_0e^{-\lambda_\pm(t-t')/2}[\hat F_a(t')\pm\hat F_b^\dagger (t')]dt'.
\end{equation}
Inview of Eqs.~\eqref{83} and ~\eqref{87} and rearranging we can write
\begin{eqnarray}\label{88}
\hat a(t)&=&E_+\hat a(0)+E_-\hat b^\dagger(0)\nonumber\\
&+&\frac{1}{2} \int^t_0 e^{-\lambda_+(t-t')/2}[\hat F_a(t')+\hat F_b^\dagger (t')]dt'+\frac{1}{2} \int^t_0 e^{-\lambda_-(t-t')/2}[\hat F_a(t')-\hat F_b^\dagger (t')]dt',
\end{eqnarray}
\begin{eqnarray}\label{89}
\hat b(t)&=&E_+ \hat b(0)+E_-\hat a^\dagger(0)\nonumber\\
&+&\frac{1}{2} \int^t_0 e^{-\lambda_+(t-t')/2}[\hat F^\dagger_a(t')+\hat F_b(t')]dt'-\frac{1}{2} \int^t_0 e^{-\lambda_-(t-t')/2}[\hat F^\dagger_a(t')-\hat F_b(t')]dt',
\end{eqnarray}
with
\begin {equation}\label{90}
E_\pm=\frac{1}{2}\bigg\lgroup e^{-\lambda_+ t/2}\pm e^{-\lambda_- t/2}\bigg\rgroup.
\end {equation}
We can separately rewrite Eq.~\eqref{88} as
\begin {equation}\label{91}
\hat a(t)=\hat a_+(t)+\hat a_-(t).
\end {equation}
One can easily show for
\begin {equation}\label{92}
\frac{d}{dt}\langle \hat a_+(t)\rangle=-\frac {\lambda_+}{2}\langle \hat a_+(t)\rangle,
\end {equation}
\begin {equation}\label{93}
\frac{d}{dt}\langle \hat a_-(t)\rangle=-\frac {\lambda_-}{2}\langle \hat a_-(t)\rangle.
\end {equation}
Inview of Eqs.~\eqref{88} and ~\eqref{91} and assuming that the cavity is initially in a two-mode vacuum state we observe that
\begin {equation}\label{94}
\langle \hat a_+(t)\rangle=\langle \hat a_-(t)\rangle=0
\end {equation}
On account of the above three equations we saw that $\hat a_+(t)$ and $\hat a_-(t)$ are Gaussian operators with vanishing means. Hence, based on Eq.~\eqref{91} $\hat a(t)$ is a Gaussian operator with zero mean. Similarily one can also show for $\hat b(t)$ to be a Gaussian operator with vanishing mean.
\par
 We also see that the formal solution of Eqs.~\eqref{81} and ~\eqref{82} is expressible as
\begin{equation}\label{95}
\hat a(t)=\hat a(0)e^{-\kappa t/2}+\int^t_0e^{-\kappa(t-t')/2}[\hat F_a(t')-\gamma\hat b^\dagger(t')]dt',
\end{equation}
\begin{equation}\label{96}
\hat b(t)=\hat b(0)e^{-\kappa t/2}+\int^t_0e^{-\kappa(t-t')/2}[\hat F_b(t')-\gamma\hat a^\dagger(t')]dt'.
\end{equation}
Now let's proceed to find a Q-function for the two-mode Subharmonic cavity light which is generally expressible as 
\begin{equation}\label{97}
Q(\alpha,\beta, t)=\frac{1}{\pi^4}\int d^2zd^2\eta \phi_a(z,\eta, t)exp(z^*\alpha+\eta^*\beta-z\alpha^*-\eta\beta^*),
\end{equation}
where, $\phi_a(z,\eta,t)$ is a characterstic function described by 
\begin{equation}\label{98}
\phi_a(z,\eta,t)=Tr(\hat \rho(0)e^{-z^*\hat a(t)}e^{z\hat a^\dagger(t)}e^{-\eta^*\hat b(t)}e^{\eta\hat b^\dagger(t)}).
\end{equation}
Making use of an identity given by Eq.~\eqref{32} along with a relation valid for Gaussian operator given in Eq.~\eqref{34} and considering the cavity radiation to be initially in a two-mode vacuum state, the characterstic function reduces to
\begin{eqnarray}\label{99}
\phi_a(z,\eta,t)&=&exp[- z^*z-\eta^*\eta] exp[-z^*z\langle \hat a^{\dagger}(t)\hat a(t)\rangle-\eta^*\eta\langle \hat b^{\dagger}(t)\hat b(t)\rangle\nonumber\\
&+&\frac{1}{2}z\eta\langle (\hat a^{\dagger}(t)\hat b^{\dagger}(t)+\hat b^{\dagger}(t)\hat a^{\dagger}(t))\rangle+\frac{1}{2}z^*\eta^*\langle (\hat a(t)\hat b(t)+\hat b(t)\hat a(t))\rangle\nonumber\\
&-&\frac{1}{2}z\eta^*\langle (\hat a^{\dagger}(t)\hat b(t)+\hat b(t)\hat a^{\dagger}(t))\rangle-\frac{1}{2}z^*\eta\langle (\hat a(t)\hat b^{\dagger}(t)+\hat b^{\dagger}(t)\hat a^{}(t))\rangle\nonumber\\
&+&\frac{1}{2}z^2\langle\hat a^{\dagger 2}(t)\rangle+\frac{1}{2}z^{*2}\langle\hat a^{2}(t)\rangle+\frac{1}{2}\eta^2\langle\hat b^{\dagger 2}(t)\rangle+\frac{1}{2}\eta^{*2}\langle\hat b^{2}(t)\rangle].                                                                   
\end{eqnarray}
Employing a relation described in Eq.~\eqref{37} along with Eq.~\eqref{81} and its conjugate we obtain
\begin {equation}\label{100}
\frac {d}{dt}\langle \hat a^\dagger(t)\hat a(t)\rangle=-\kappa\langle \hat a^\dagger(t)\hat a(t)\rangle-\gamma(\langle \hat a^\dagger(t)\hat b^\dagger(t)\rangle+\langle \hat a(t)\hat b(t)\rangle)+\langle \hat F_a^\dagger(t)\hat a(t)\rangle+\langle \hat a^\dagger \hat F_a(t),
\end {equation}
compairing with Eq.~\eqref{76} we see that
\begin {equation}\label{101}
\langle \hat F_a^\dagger(t)\hat a(t)\rangle+\langle \hat a^\dagger(t) \hat F_a(t)=0,
\end {equation}
from which we have
\begin {equation}\label{102}
\langle \hat F_a^\dagger(t')\hat F_a(t)\rangle=0.
\end {equation}
Similarly, one can also show for 
\begin {equation}\label{103}
\langle \hat F_b^\dagger(t')\hat F_b(t)\rangle=\langle \hat F_b^\dagger(t')\hat F_a(t)\rangle=\langle \hat F_a^\dagger(t')\hat F_b(t)\rangle=\langle \hat F_a^\dagger(t')\hat F_b^\dagger(t)\rangle=\langle \hat F_b(t')\hat F_a(t)\rangle=0.
\end {equation}
Using a relation given in Eq.~\eqref{54} along with Eq.~\eqref{81} and its conjugate together with a commutation relation described in Eq.~\eqref{5} we have
\begin {equation}\label{104}
\frac {d}{dt}\langle \hat a^\dagger(t)\hat a(t)\rangle=-\kappa\langle \hat a^\dagger(t)\hat a(t)\rangle-\gamma(\langle \hat a^\dagger(t)\hat b^\dagger(t)\rangle+\langle \hat a(t)\hat b(t)\rangle)+\langle \hat F_a(t)\hat a^\dagger(t)\rangle+\langle \hat a(t) \hat F^\dagger_a(t)-\kappa,
\end {equation}
compairing with Eq.~\eqref{76} one can observe that
\begin {equation}\label{105}
\langle \hat F_a(t)\hat a^\dagger(t)\rangle+\langle \hat a (t) \hat F^\dagger_a(t)=\kappa,
\end {equation}
which yields
\begin {equation}\label{106}
\langle \hat F_a(t')\hat F^\dagger_a(t'')\rangle=\kappa\delta (t'-t'').
\end {equation}
Using Eqs.~\eqref{88} and ~\eqref{89} and its conjugates, at steady state, we obtain
\begin {equation}\label{107}
\langle \hat a^\dagger(t)\hat a(t)\rangle=\langle \hat b^\dagger(t)\hat b(t)\rangle=\frac {2\gamma^2}{\kappa^2-4\gamma^2},
\end {equation}
\begin {equation}\label{108}
\langle \hat a^\dagger(t)\hat b^\dagger(t)\rangle=\langle \hat a(t)\hat b(t)\rangle=-\frac {\kappa\gamma}{\kappa^2-4\gamma^2}.
\end {equation}
One can also check for 
\begin{eqnarray}\label{109}
\langle \hat a^{\dagger}(t)\hat b(t)\rangle=\langle \hat a(t)\hat b^{\dagger}(t)\rangle=0,                                                              
\end{eqnarray}
\begin {equation}\label{110}
\langle\hat a^{\dagger 2}(t)\rangle=\langle\hat a^{2}(t)\rangle=\langle\hat b^{\dagger 2}(t)\rangle=\langle\hat b^{2}(t)\rangle=0. 
\end {equation}
Inview of the above results the characterstic function reduces to
\begin{eqnarray}\label{111}
\phi_a(z,\eta,t)=exp[-a(z^*z+\eta^*\eta)-b(z\eta+z^*\eta^*)],                                                     
\end{eqnarray}
where,
\begin{equation}\label{112}
a=\frac{\kappa^2-2\gamma^2}{\kappa^2-4\gamma^2},
\end{equation}
and 
\begin{equation}\label{113}
b=\frac{\kappa\gamma}{\kappa^2-4\gamma^2}.
\end{equation}
Upon substituting Eq.~\eqref{111} in Eq.~\eqref{97} we see that
\begin{equation}\label{114}
Q(\alpha,\beta)=\frac{1}{\pi^2}\int\frac{d^2zd^2\eta}{\pi^2}exp(-az^*z-a\eta^*\eta-bz\eta-bz^*\eta^*+z^*\alpha+\eta^*\beta-z\alpha^*-\eta\beta^*),
\end{equation}
performing the integration the Q-function for a two-mode Subharmonic cavity light takes the form 
\begin{equation}\label{115}
Q(\alpha,\beta)=\frac{u^2-v^2}{\pi^2}exp[-u(\alpha^*\alpha+\beta^*\beta)-v(\alpha\beta+\alpha^*\beta^*)],
\end{equation}
where,
\begin{equation}\label{116}
u=\frac{a}{a^2-b^2},
\end{equation}
\begin{equation}\label{117}
v=\frac{b}{a^2-b^2}.
\end{equation}
\section {A Superposed Two-Mode Coherent and Subharmonic Light}
Here we first seek to obtain Q-function for a superposed two-mode coherent and subharmonic light. Q-function for the Superposed two-mode lights is generally expressible as
\begin{eqnarray}\label{118}
Q(\alpha,\beta)&=&\int\frac{d^2\chi d^2\lambda d^2\eta d^2\gamma}{\pi^2}Q(\chi^*,\lambda^*,\alpha-\gamma,\beta-\eta)Q(\gamma^*,\eta^*,\alpha-\chi,\beta-\lambda)\nonumber\\
&&exp[-|\alpha-\chi-\gamma|^2-|\beta-\lambda-\eta|^2].
\end{eqnarray}
We now proceed to find Q-function for the Superposed two-mode coherent and Subharmonic light beams. Thus, inview of Eqs.~\eqref{69} and ~\eqref{115}, we have
\begin{eqnarray}\label{119}
Q(\chi^*,\lambda^*,\alpha-\gamma,\beta-\eta)&=&\frac{1}{\pi^2}exp[-\chi^*\alpha+\chi^*\gamma-\lambda^*\beta+\lambda^*\eta\nonumber\\
&+&(\alpha-\gamma+\chi^*+\beta-\eta+\lambda^*)q-2q^2],
\end{eqnarray}
and
\begin{eqnarray}\label{120}
Q(\gamma^*,\eta^*,\alpha-\chi,\beta-\lambda)&=&\frac{u^2-v^2}{\pi^2}exp[-u\gamma^*\alpha+u\gamma^*\chi-u\eta^*\beta+u\eta^*\lambda\nonumber\\
&-&v\alpha\beta+v\alpha\lambda+v\chi\beta-v\chi\lambda-v\gamma^*\eta^*].
\end{eqnarray}
Substituting Eqs.~\eqref{119} and ~\eqref{120} in Eq.~\eqref{118} the Q-function for the Superposed two-mode coherent and Subharmonic light beams become
\begin{eqnarray}\label{121}
Q(\alpha,\beta,t)&=&\frac{u^2-v^2}{\pi^2}exp[-u(\alpha^*\alpha+\beta^*\beta)-v(\alpha\beta+\alpha^*\beta^*)\nonumber\\
&+&q(u+v)(\alpha+\alpha^*+\beta+\beta^*)-2q^2(u+v)].
\end{eqnarray}
\section{Mean Photon Number}
Mean photon number of a superposed two-mode light is defined as
\begin {equation}\label{122}
\bar n=\langle \hat c^\dagger\hat c\rangle,
\end {equation}
where,
\begin {equation}\label{123}
\hat c=\hat a+\hat b,
\end {equation}
is annihilation operator for a superposed two-mode light. Thus, inview of Eq.~\eqref{123} the mean photon number takes a form
\begin {equation}\label{124}
\bar n=\langle \hat a^\dagger\hat a\rangle+\langle \hat a^\dagger \hat b\rangle+\langle \hat b^\dagger \hat a\rangle+\langle \hat b^\dagger\hat b\rangle.
\end {equation}
Using Q-function in the antinormal order we can evaluate
\begin {equation}\label{125}
\langle \hat a^\dagger\hat a\rangle=\int d^2\alpha Q(\alpha^*,\alpha)\alpha^*\alpha-1,
\end {equation}
with,
\begin{equation}\label{126}
Q(\alpha^*,\alpha)=\int d^2\beta Q(\alpha,\beta),
\end{equation}
substituting Eq.~\eqref{121} in ~\eqref{126} the Q-function for mode-a takes a form
\begin {eqnarray}\label{127}
Q(\alpha^*,\alpha)&=&\frac{u^2-v^2}{\pi u}exp\bigg[-\bigg(\frac {u^2-v^2}{u}\bigg)\alpha^*\alpha+q\bigg(\frac {u^2-v^2}{u}\bigg)\alpha^*\nonumber\\
&+&q\bigg(\frac {u^2-v^2}{u}\bigg)\alpha-q^2\bigg(\frac {u^2-v^2}{u}\bigg)\bigg],
\end {eqnarray}
upon replacing the above Q-function in Eq.~\eqref{125} and rearranging we see that
\begin {eqnarray}\label{128}
\langle \hat a^\dagger\hat a\rangle&=&\frac {u}{\pi q^2(u^2-v^2)}\frac {\partial}{\partial a\partial b}exp\bigg[-\bigg(\frac {u^2-v^2}{u}\bigg)\bigg]\int d^2\alpha exp\bigg[-\bigg(\frac {u^2-v^2}{u}\bigg)\alpha^*\alpha\nonumber\\
&+&qa\bigg(\frac {u^2-v^2}{u}\bigg)\alpha^*+qb\bigg(\frac {u^2-v^2}{u}\bigg)\alpha\bigg]\bigg|_{a=b=1}-1.
\end {eqnarray}
Performing the integration and partial differentiation and setting $a=b=1$ yields
\begin {eqnarray}\label{129}
\langle \hat a^\dagger\hat a\rangle=\frac {u}{u^2-v^2}+q^2-1.
\end {eqnarray}
Similarly,
\begin {eqnarray}\label{130}
\langle \hat b^\dagger\hat b\rangle=\frac {u}{u^2-v^2}+q^2-1.
\end {eqnarray}
Inview of the Q-function described in Eq.~\eqref{121} we can also evaluate and obtain
\begin {eqnarray}\label{131}
\langle \hat a^\dagger\hat b\rangle=\langle \hat b^\dagger \hat a\rangle=q^2.
\end {eqnarray}
Finally upon substituting Eqs.~\eqref{129}, ~\eqref{130} and ~\eqref{131} into Eq.~\eqref{124} and on account of Eqs.~\eqref{21},~\eqref{112}, ~\eqref{113}, ~\eqref{116} and ~\eqref{117} and rearranging
\begin{equation}\label{132}
\bar n=\frac{4\gamma^2}{\kappa^2-4\gamma^2}+\frac{16\varepsilon^2}{\kappa^2}(1-e^{-\kappa t/2})^2,
\end{equation}
at steady state the mean photon number for a superposed two-mode coherent and Subharmonic cavity light takes the form
\begin{equation}\label{133}
\bar n_{ss}=\frac{4\gamma^2}{\kappa^2-4\gamma^2}+\frac{16\varepsilon^2}{\kappa^2}.
\end{equation}
\section{Variance of a Photon Number}
In this section we seek to calculate the variance of the photon number for a superposed two-mode coherent and subharmonic cavity light. To this end, we define variance of a photon number as
\begin{equation}\label{134}
(\Delta n)^2=\langle \hat n^2\rangle-\langle \hat n\rangle^2.
\end{equation}
Where photon number for a superposed two-mode light is 
\begin{equation}\label{135}
\hat n=\hat c^\dagger\hat c,
\end{equation}
Using the above equation variance of the photon number becomes
\begin{equation}\label{136}
(\Delta n)^2=\langle \hat c^\dagger\hat c\hat c^\dagger\hat c\rangle-\langle \hat c^\dagger\hat c\rangle^2.
\end{equation}
Inview of a commutation relation
\begin{equation}\label{137}
[\hat c,\hat c^\dagger]=4,
\end{equation}
the variance of the photon number takes a form
\begin{equation}\label{138}
(\Delta n)^2=\langle \hat c^{\dagger2}\hat c^{2}\rangle+4\bar n-\bar n^2.
\end{equation}
Inview of Eqs.~\eqref{23} and ~\eqref{24}  we can write 
\begin{equation}\label{139}
\hat a=q+\hat a',
\end{equation}
\begin{equation}\label{140}
\hat b=q+\hat b',
\end{equation}
with,
\begin{equation}\label{141}
\hat a'=\hat a_1'+\hat a_2,
\end{equation}
and 
\begin{equation}\label{142}
\hat b'=\hat b'_1+\hat b_2.
\end{equation}
where in this case $\hat a_1,\hat a_2$ are annihilation operators for mode-a of coherent and subharmonic lights respectively. We also know from Eqs.~\eqref{27},~\eqref{28} and ~\eqref{29} that $\hat a'_1$ and $\hat b'_1$ are Gaussian operators with vanishing means. And Inview of Eqs.~\eqref{88},~\eqref{89},~\eqref{91},~\eqref{92},~\eqref{93} and ~\eqref{94} and assuming the cavity radiation to be initially in a two-mode vacuum state we see that  $\hat a_2$ and $\hat b_2$ are Gaussian operators with vanishing mean. 
\par
Thus, from the above two equations we see that $\hat a'$ and $\hat b'$ are Gaussian operators with zero mean, because it is the sum of two Gaussian operators with zero mean. Hence, using Eqs.~\eqref{139} and ~\eqref{140} we have
\begin{equation}\label{143}
\hat c=2q+\hat c',
\end{equation}
in this case
\begin{equation}\label{144}
\hat c'=\hat a'+\hat b'.
\end{equation}
From the above equation one can easily see that $c'(t)$ is also a Gaussian operator with zero mean. Hence, inview Eq.~\eqref{143} we have
\begin{eqnarray}\label{145}
\langle \hat c^{\dagger2}\hat c^{2}\rangle&=&\langle (2q+\hat c'^{\dagger})^2(2q+\hat c')^{2}\rangle\nonumber\\
&=&\langle \hat c'^{\dagger2}\hat c'^2\rangle+4q\langle \hat c'^{\dagger2}\hat c'\rangle+4q^2\langle \hat c'^{\dagger2}\rangle+4q\langle \hat c'^{\dagger}\hat c'^2\rangle+16q^2\langle \hat c'^{\dagger}\hat c'\rangle\nonumber\\
&+&16q^3\langle \hat c'^{\dagger}\rangle+4q^2\langle \hat c'^2\rangle+16q^3\langle \hat c'\rangle+16q^4,
\end{eqnarray}
using the fact that $\hat c'$ is a Gaussian operator with zero mean and the expectation value of odd number of Gaussian operators is zero Eq.~\eqref{145} reduces to
\begin{eqnarray}\label{146}
\langle \hat c^{\dagger2}\hat c^{2}\rangle=\langle \hat c'^{\dagger2}\hat c'^2\rangle+4q^2\langle \hat c'^{\dagger2}\rangle+16q^2\langle \hat c'^{\dagger}\hat c'\rangle+4q^2\langle \hat c'^2\rangle+16q^4.
\end{eqnarray}
Employing a relation, valid for Gaussian operators[1]
\begin{equation}\label{147}
\langle \hat A\hat B\hat C \hat D\rangle=\langle \hat A\hat B\rangle\langle \hat C \hat D\rangle+\langle \hat A\hat C \rangle\langle \hat B \hat D\rangle+\langle \hat A \hat D\rangle\langle \hat B\hat C \rangle,
\end{equation}
Eq.~\eqref{146} takes a form
\begin{eqnarray}\label{148}
\langle \hat c^{\dagger2}\hat c^{2}\rangle=\langle \hat c'^{\dagger2}\rangle\langle\hat c'^2\rangle+2\langle\hat c'^{\dagger}\hat c'\rangle^2+4q^2\langle \hat c'^{\dagger2}\rangle+16q^2\langle \hat c'^{\dagger}\hat c'\rangle+4q^2\langle \hat c'^2\rangle+16q^4.
\end{eqnarray}
Using Eqs.~\eqref{141}, ~\eqref{142}, ~\eqref{144} and also inview of Eqs.~\eqref{58} upto ~\eqref{65} along with Eqs.~\eqref{51} and ~\eqref{52}, we obtain
\begin{eqnarray}\label{149}
\langle \hat c'^2\rangle=2\langle \hat a_2\hat b_2\rangle.
\end{eqnarray}
Similarily,
\begin{eqnarray}\label{150}
\langle \hat c'^{\dagger2}\rangle=2\langle \hat a^\dagger_2\hat b^\dagger_2\rangle,
\end{eqnarray}
and
\begin{eqnarray}\label{151}
\langle \hat c'^{\dagger}\hat c'\rangle= 2 \langle \hat a^{\dagger}_2\hat a_2\rangle.
\end{eqnarray}
On account of the above three results we can write Eq.~\eqref{148} as
\begin{eqnarray}\label{152}
\langle \hat c^{\dagger2}\hat c^{2}\rangle=4\langle \hat a_2\hat b_2\rangle^2+8\langle \hat a^{\dagger}_2\hat a_2\rangle^2+16q^2\langle \hat a_2\hat b_2\rangle+32q^2\langle \hat a^\dagger_2\hat a_2\rangle+16q^4.
\end{eqnarray}
Inserting Eq.~\eqref{152} into Eq.~\eqref{138} the variance of the photon number becomes
\begin {eqnarray}\label{153}
(\Delta n)^2=4\langle \hat a_2\hat b_2\rangle^2+8\langle \hat a^{\dagger}_2\hat a_2\rangle^2+16q^2\langle \hat a_2\hat b_2\rangle+32q^2\langle \hat a^\dagger_2\hat a_2\rangle+16q^4+4\bar n-\bar n^2,
\end {eqnarray}
where $\bar n$ is the mean photon number of a superposed two-mode coherent and subharmonic cavity light. 
Now let's evaluate the expectation values in the above equation 
\begin{equation}\label{154}
\langle \hat a_2\hat b_2\rangle=\int d^2\alpha_2 d^2\beta_2 Q(\alpha_2,\beta_2)\alpha_2\beta_2,
\end{equation}
the Q-function described in Eq.~\eqref{115} expressible in the antinormal order as
\begin {eqnarray}\label{155}
Q(\alpha_2,\beta_2)=\frac{u^2-v^2}{\pi^2}exp[-u\alpha_2^*\alpha_2-u\beta_2^*\beta_2-v\alpha_2\beta_2-v\alpha_2^*\beta_2^*].
\end {eqnarray}
Substituting the above Q-function into Eq.~\eqref{154}, rearranging and Integrating we have
\begin{eqnarray}\label{156}
\langle \hat a_2\hat b_2\rangle=-\frac{v}{u^2-v^2}.
\end{eqnarray}
Finally making use of Eqs.~\eqref{112}, ~\eqref{113}, ~\eqref{116} and ~\eqref{117} yields
\begin{eqnarray}\label{157}
\langle \hat a_2\hat b_2\rangle=-\frac{\kappa\gamma}{\kappa^2-4\gamma^2}.
\end{eqnarray}
We also have
\begin{equation}\label{158}
\langle \hat a^\dagger_2\hat a_2\rangle=\int d^2\alpha_2Q(\alpha^*_2,\alpha_2)\alpha_2^*\alpha_2-1.
\end{equation}
Integrating Eq.~\eqref{115} with respect to $\beta$ we have a Q-function in the antinormal order as
\begin{equation}\label{159}
Q(\alpha^*_2,\alpha_2)=\frac{u^2-v^2}{\pi u}exp\bigg[-\bigg(\frac{u^2-v^2}{u}\bigg)\alpha^*_2\alpha_2\bigg],
\end{equation}
substituting the above Q-function rearranging and performing the integration yields
\begin{equation}\label{160}
\langle \hat a^\dagger_2\hat a_2\rangle=\bigg[\frac{u^2-v^2}{u}-1\bigg].
\end{equation}
Inview of Eqs.~\eqref{116}, ~\eqref{117} the above equation takes a form
\begin{equation}\label{161}
\langle \hat a^\dagger_2\hat a_2\rangle=a-1.
\end{equation}
Finally using Eq.~\eqref{112} we obtain
\begin{equation}\label{162}
\langle \hat a^\dagger_2\hat a_2\rangle=\frac{2\gamma^2}{\kappa^2-4\gamma^2}.
\end{equation}
Thus, substituting Eqs.~\eqref{21}, ~\eqref{157} and ~\eqref{162} into Eq.~\eqref{153} variance of a photon number for a superposed two-mode coherent and subharmonic cavity light becomes
\begin {eqnarray}\label{163}
(\Delta n)^2&=&\frac{16\gamma^4}{(\kappa^2-4\gamma^2)^2}+\frac{4\kappa^2\gamma^2}{(\kappa^2-4\gamma^2)^2}+\frac{16\gamma^2}{(\kappa^2-4\gamma^2)}\nonumber\\
&+&\frac{128\varepsilon^2\gamma^2}{\kappa^2(\kappa^2-4\gamma^2)}(1-e^{-\kappa t/2})^2-\frac{64\varepsilon^2\gamma}{\kappa(\kappa^2-4\gamma^2)}(1-e^{-\kappa t/2})^2\nonumber\\
&+&\frac{64\varepsilon^2}{\kappa^2}(1-e^{-\kappa t/2})^2,
\end {eqnarray}
at steady state it takes a form
\begin {eqnarray}\label{164}
(\Delta n)^2&=&\frac{16\gamma^4}{(\kappa^2-4\gamma^2)^2}+\frac{4\kappa^2\gamma^2}{(\kappa^2-4\gamma^2)^2}+\frac{16\gamma^2}{(\kappa^2-4\gamma^2)}\nonumber\\
&+&\frac{128\varepsilon^2\gamma^2}{\kappa^2(\kappa^2-4\gamma^2)}-\frac{64\varepsilon^2\gamma}{\kappa(\kappa^2-4\gamma^2)}+\frac{64\varepsilon^2}{\kappa^2}.
\end {eqnarray}
\section{Quadrature Variance}
We now wish to calculate the quadrature variance of a superposed two-mode coherent and subharmonic cavity light beams. To this end, we define the plus and minus quadrature operators for a two-mode light as
\begin{equation}\label{165}
\hat c_+=\hat c^\dagger+\hat c,
\end{equation}
and 
\begin{equation}\label{166}
\hat c_-=i(\hat c^\dagger-\hat c).
\end{equation}
Variance of plus and minus quadrature operators is giben by
\begin{eqnarray}\label{167}
(\Delta c_\pm)^2&=&\langle \hat c_\pm,\hat c_\pm\rangle\nonumber\\
&=&\langle \hat c_\pm^2\rangle-\langle \hat c_\pm\rangle^2.
\end{eqnarray}
Inview of Eqs.~\eqref{165} and ~\eqref{166} and the commutation relation given by Eq.~\eqref{137} the quadrature variance becomes
\begin{eqnarray}\label{168}
(\Delta c_\pm)^2&=&4+2\langle \hat c^\dagger\hat c\rangle\pm\langle \hat c^{\dagger2}\rangle\pm\langle\hat c^2\rangle\mp\langle \hat c^{\dagger}\rangle^2\mp\langle \hat c\rangle^2-2\langle \hat c^{\dagger}\rangle\langle \hat c\rangle,
\end{eqnarray}
from which the variance of the plus and minus quadrature operators can be written separately as
\begin{eqnarray}\label{169}
(\Delta c_+)^2&=&4+2\langle \hat c^\dagger\hat c\rangle+\langle \hat c^{\dagger2}\rangle+\langle\hat c^2\rangle-\langle \hat c^{\dagger}\rangle^2-\langle \hat c\rangle^2-2\langle \hat c^{\dagger}\rangle\langle \hat c\rangle,
\end{eqnarray}
and 
\begin{eqnarray}\label{170}
(\Delta c_-)^2&=&4+2\langle \hat c^\dagger\hat c\rangle-\langle \hat c^{\dagger2}\rangle-\langle\hat c^2\rangle+\langle \hat c^{\dagger}\rangle^2+\langle \hat c\rangle^2-2\langle \hat c^{\dagger}\rangle\langle \hat c\rangle.
\end{eqnarray}
Using Eq.~\eqref{143} and based on the fact that $\hat c'$ is a Gaussian operator with vanishing mean the above two equations takes a form
\begin{eqnarray}\label{171}
(\Delta c_+)^2=4+2\langle \hat c'^{\dagger}\hat c'\rangle+\langle \hat c'^{\dagger2}\rangle+\langle\hat c'^2\rangle,
\end{eqnarray}
and 
\begin{eqnarray}\label{172}
(\Delta c_-)^2=4+2\langle \hat c'^{\dagger}\hat c'\rangle-\langle \hat c'^{\dagger2}\rangle-\langle\hat c'^2\rangle.
\end{eqnarray}
Inview of Eqs.~\eqref{149} to ~\eqref{151} the above two equations becomes
\begin{eqnarray}\label{173}
(\Delta c_+)^2=4+4\langle \hat a^{\dagger}_2\hat a_2\rangle+4\langle \hat a_2\hat b_2\rangle,
\end{eqnarray}
\begin{eqnarray}\label{174}
(\Delta c_-)^2=4+4\langle \hat a^{\dagger}_2\hat a_2\rangle-4\langle \hat a_2\hat b_2\rangle.
\end{eqnarray}
Finally on account of Eqs.~\eqref{157} and ~\eqref{162} and rearranging, the plus and minus quadrature variance for a superposed two-mode coherent and subharmonic cavity light takes a form
\begin{eqnarray}\label{175}
(\Delta c_+)^2=4-\frac{4\gamma}{\kappa+2\gamma},
\end{eqnarray}
\begin{eqnarray}\label{176}
(\Delta c_-)^2=4+\frac{4\gamma}{\kappa-2\gamma}.
\end{eqnarray}
Generally, 
\begin{equation}\label{177}
(\Delta c_\pm)^2=4\mp\frac{4\gamma}{\kappa\pm2\gamma}.
\end{equation}
\begin{flushleft}
\begin {figure}[h]
\includegraphics [height=10cm, width =14cm ]{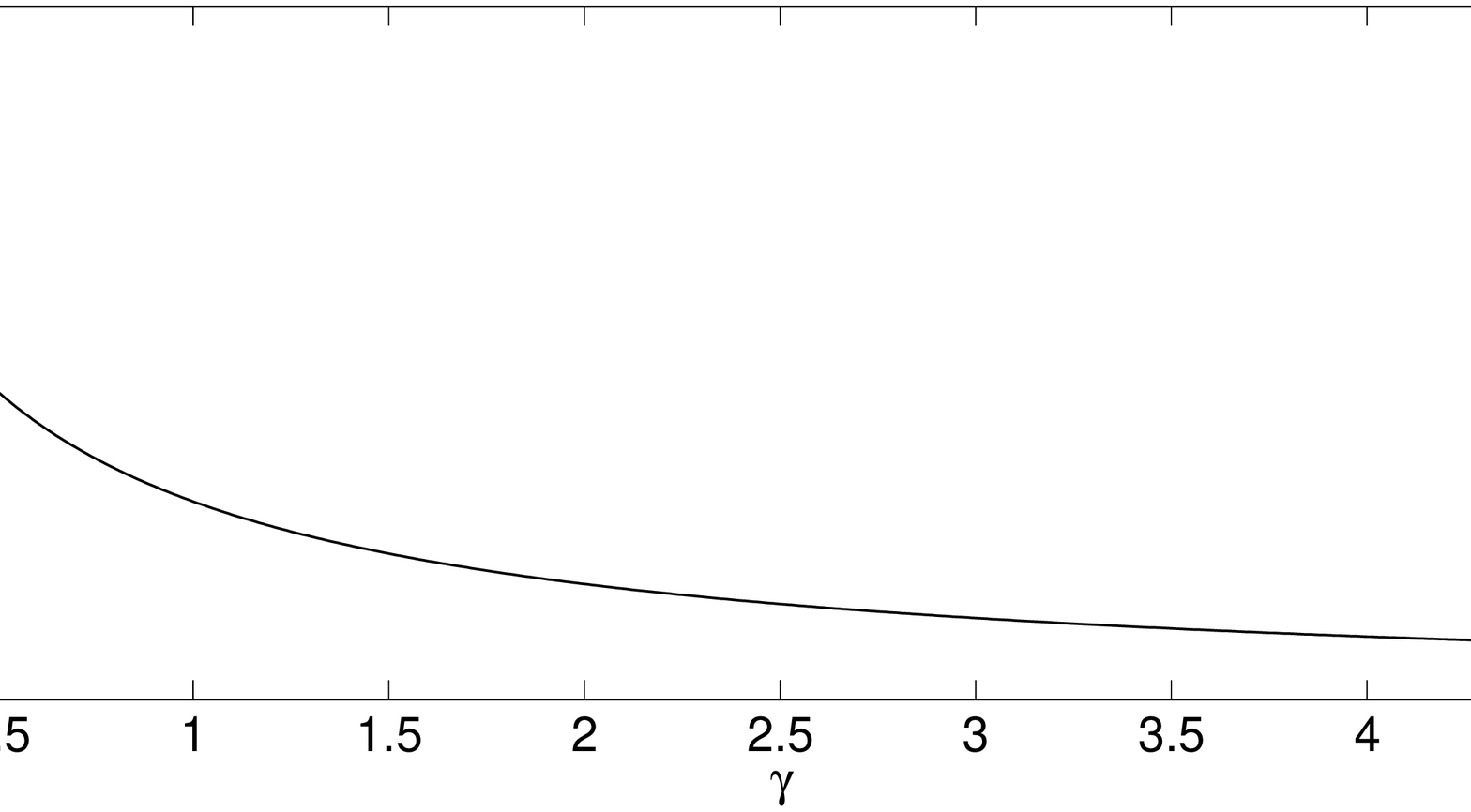}
\caption{ A plot of plus quadrature Variance versus $\gamma$ for $\kappa=0.8$ }
\label {fig:p}
\end {figure}
\end {flushleft}
\par
From the above equations and figure below one can clearly see that the superposed two-mode coherent and subharmonic cavity light is in a squeezed state and the Squeezing occurs in the plus quadrature. In addition, we observe that for $\kappa=2\gamma$ the variance of the minus quadrature diverges. We then identity $\kappa=2\gamma$ as the threshold condition.
\newpage 
\section{Quadrature Squeezing}
We now proceed to determine the quadrature Squeezing of a superposed two-mode coherent and subharmonic cavity light. To this end upon setting $\gamma=0$ into Eqs.~\eqref{175} and ~\eqref{176} we see that  
\begin{eqnarray}\label{178}
(\Delta c_\pm)_v^2=4.
\end{eqnarray}
So we observe for $\gamma=0$ that a superposed two-mode coherent and subharmonic cavity light is in a two-mode vacuum state in which the uncertainities in the two quadratures are equal and satisfies minimum uncertainity relation.
\par
Now let's determine the quadrature Squeezing of a superposed two-mode coherent and subharmonic cavity light relative to the quadrature variance of the two-mode cavity vacuum state. We define quadrature Squeezing of a two-mode cavity light by
\begin{equation}\label{179}
S=\frac{(\Delta c_\pm)_v^2-(\Delta c_\pm)^2}{(\Delta c_\pm)_v^2},
\end{equation}
substituting Eqs.~\eqref{175} and ~\eqref{178} into Eq.~\eqref{179} and rearranging quadrature Squeezing for a superposed two-mode cavity light takes a form
\begin{equation}\label{180}
S=\frac{\gamma}{\kappa+2\gamma}.
\end{equation}
The above equation represents the global quadrature Squeezing of a superposed two-mode coherent and subharmonic cavity light. We can easily observe that at steady state and threshold there is a $25\%$ quadrature Squeezing below the vacuum state level.
\begin{center}
\begin {figure}[h]
\centering
\includegraphics [height=10cm, width =14cm ]{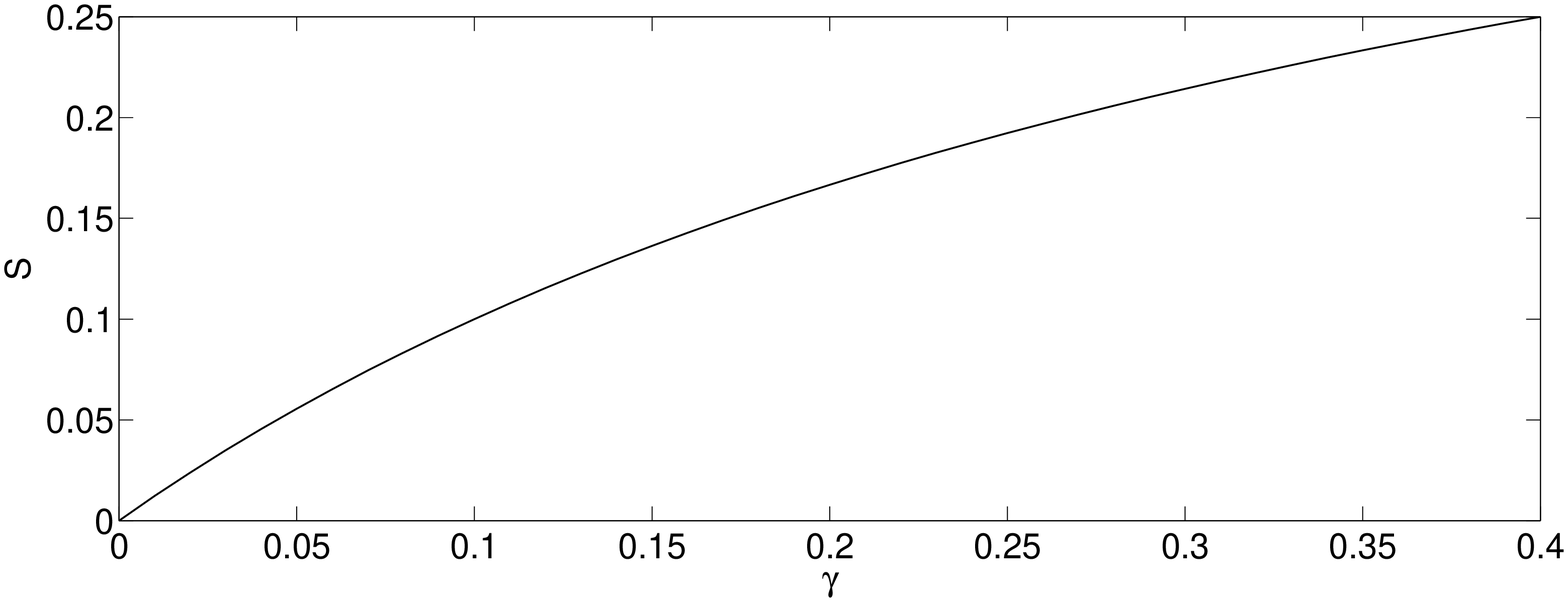}
\caption{ A plot of quadrature squeezing versus $\gamma$ for $\kappa=0.8$ }
\label {fig:p}
\end {figure}
\end {center}
\section{Cauchy-Schwarz Inequality }
In this section we use the second-order correlation to determine the entanglement of a system with two-modes, $\hat a$ and $\hat b$. The system is said to be entangled if it satisfies the Cauchy-Schwarz equality. The expectation value of cross-correlation between two-modes $\langle \hat a^\dagger\hat a\hat b^\dagger\hat b\rangle$ is bounded by [6,7]
\begin{equation}\label{181}
\langle\hat a^{\dagger2}\hat a^2\rangle\langle\hat b^{\dagger2}\hat b^2\rangle\geq\langle \hat a^\dagger\hat a\hat b^\dagger\hat b\rangle^2.
\end{equation}
The inequality may be expressed in terms of the second-order correlation functions at zero times as
\begin{equation}\label{182}
g^{(2)}_a(0)=\frac{\langle\hat a^{\dagger2}\hat a^2\rangle}{\langle\hat a^{\dagger}\hat a\rangle^2},
\end{equation}
and 
\begin{equation}\label{183}
g^{(2)}_b(0)=\frac{\langle\hat b^{\dagger2}\hat b^2\rangle}{\langle\hat b^{\dagger}\hat b\rangle^2}.
\end{equation}
The second-order cross-correlation function is also expressible as
\begin{equation}\label{184}
g^{(2)}_{ab}(0)=\frac{\langle\hat b^{\dagger}\hat a^\dagger\hat a\hat b\rangle}{\langle\hat a^{\dagger}\hat a\rangle\langle\hat b^{\dagger}\hat b\rangle}.
\end{equation}
Assuming that $\langle\hat b^{\dagger}\hat a^\dagger\hat a\hat b\rangle=\langle\hat a^\dagger\hat a\hat b^{\dagger}\hat b\rangle$, the Cauchy-Schwarz inequality can be expressed as
\begin{equation}\label{185}
g^{(2)}_a(0)g^{(2)}_b(0)\geq[g^{(2)}_{ab}(0)]^2.
\end{equation}
The entanglement in a system can be identified if the above inequality is satisfied. Hence, we need to show for the above inequality for a superposed two-mode coherent and subharmonic cavity light. 
\newline
Employing Eqs.~\eqref{23} and ~\eqref{24} we can rewrite Eqs.~\eqref{182} and ~\eqref{183} as
\begin{equation}\label{186}
g^{(2)}_a(0)=\frac{\langle(q+\hat a'^{\dagger})^2(q+\hat a')^2\rangle}{\langle(q+\hat a'^{\dagger})(q+\hat a')\rangle^2}.
\end{equation}
We know that $\hat a'$ is Gaussian operator with zero mean and expectation value of odd number of a Gaussian operators is zero, thus we have  
\begin {eqnarray}\label{187}
g^{(2)}_a(0)&=&\frac{q^4+4q^2\langle\hat a_2^{\dagger}\hat a_2\rangle+2\langle\hat \hat a_2^{\dagger}\hat a_2\rangle^2}{(q^2+\langle\hat a_2^{\dagger}\hat a_2\rangle)^2}\nonumber\\
&=&1+\frac{\langle\hat \hat a_2^{\dagger}\hat a_2\rangle^2+2q^2\langle\hat a_2^{\dagger}\hat a_2\rangle}{(q^2+\langle\hat a_2^{\dagger}\hat a_2\rangle)^2},
\end {eqnarray}
at steady state and inview of Eqs.~\eqref{21} and ~\eqref{162} we observe that
\begin {eqnarray}\label{188}
g^{(2)}_a(0)&=&1+\frac{\frac{4\gamma^4}{(\kappa^2-4\gamma^2)^2}+\frac{16\varepsilon^2\gamma^2}{\kappa^2(\kappa^2-4\gamma^2)}}{(\frac{4\varepsilon^2}{\kappa^2}+\frac{2\gamma^2}{\kappa^2-4\gamma^2})^2}\nonumber\\
&=&1+\frac{\kappa^2[4\gamma^4\kappa^2+16\varepsilon^2\gamma^2(\kappa^2-4\gamma^2)]}{(4\varepsilon^2\kappa^2-16\varepsilon^2\gamma^2+2\gamma^2\kappa^2)^2},
\end {eqnarray}
at threshold ($\kappa=2\gamma$) the second-order correlation function reduces to
\begin {eqnarray}\label{189}
g^{(2)}_a(0)=2.
\end {eqnarray}
Following the same procedure one can also show at steady state and threshold that  
\begin {eqnarray}\label{190}
g^{(2)}_b(0)=2.
\end {eqnarray}
We now proceed to calculate the second-order cross-correlation function. To this end, inview of Eqs.~\eqref{23} and ~\eqref{24} we rewrite Eq.~\eqref{184} as
\begin{equation}\label{191}
g^{(2)}_{ab}(0)=\frac{\langle(q+\hat a'^{\dagger})(q+\hat a')(q+\hat b'^{\dagger})(q+\hat b')\rangle}{\langle(q+\hat a'^{\dagger})(q+\hat a')\rangle\langle(q+\hat b'^{\dagger})(q+\hat b')\rangle},
\end{equation}
which follows then
\begin{eqnarray}\label{192}
g^{(2)}_{ab}(0)&=&\frac{q^4+2q^2\langle\hat a_2^\dagger\hat a_2\rangle+2q^2\langle\hat a_2\hat b_2\rangle+\langle \hat a_2^\dagger\hat a_2\rangle^2+\langle \hat a_2\hat b_2\rangle^2}{(q^2+\langle\hat a_2^\dagger\hat a_2\rangle)^2}\nonumber\\
&=&1+\frac{2q^2\langle\hat a_2\hat b_2\rangle+\langle \hat a_2\hat b_2\rangle^2}{(q^2+\langle\hat a_2^\dagger\hat a_2\rangle)^2}..
\end{eqnarray}
At steady state and using Eqs.~\eqref{21} and ~\eqref{157} we get
\begin{eqnarray}\label{193}
g^{(2)}_{ab}(0)&=&1+\frac{\frac{\kappa^2\gamma^2}{(\kappa^2-4\gamma^2)^2}-\frac{8\varepsilon^2\gamma}{\kappa(\kappa^2-4\gamma^2)}}{(\frac{4\varepsilon^2}{\kappa^2}+\frac{2\gamma^2}{\kappa^2-4\gamma^2})^2}\nonumber\\
&=&1+\frac{[\kappa^2\gamma^2-\frac{8\varepsilon^2\gamma}{\kappa}(\kappa^2-4\gamma^2)]\kappa^4}{(4\varepsilon^2\kappa^2-16\varepsilon^2\gamma^2+2\gamma^2\kappa^2)^2}.
\end{eqnarray}
Thus, at steady state and threshold ($\kappa=2\gamma$) the second-order cross-correlation function becomes
\begin{equation}\label{194}
g^{(2)}_{ab}(0)=2.
\end{equation}
Finally inview of Eqs.~\eqref{189}, ~\eqref{190} and ~\eqref{194} we see that the inequality given in Eq.~\eqref{185} is satisfied. That is $4\geq4$, so a superposed two-mode coherent and subharmonic cavity light is entangled.
\section{Continuous Variable(CV) Entanglement}
The other entanglement criterion is proposed by Simon and DGCZ, they proposed a class of condition that is sufficient to show entanglement in continuous variable system. According to this criteria, a system is known to be entangled if a quantum fluctuations of the two Einstein-Podolsky-Rosen (EPR) like operators, $\hat u$ and $\hat v$, of a superposed two-mode system satisfy the inequality [6,7]
\begin{equation}\label{195}
(\Delta u)^2+(\Delta v)^2<4.
\end{equation}
Where,
\begin{equation}\label{196}
\hat u=\hat x_1+\hat x_2,
\end{equation}
\begin{equation}\label{197}
\hat v=\hat p_1-\hat p_2,
\end{equation}
and 
\begin{equation}\label{198}
\hat x_j=\frac{\hat a_j+\hat a_j^\dagger}{\sqrt{2}},
\end{equation}
\begin{equation}\label{199}
\hat p_j=\frac{\hat a_j-\hat a_j^\dagger}{\sqrt{2}i}.
\end{equation}
with $(j=1,2)$ are the quadratures for the two-modes of the cavity field. On substituting the definitions of $\hat u$ and $\hat v$ the sum of the variance of the two EPR like operators can be written as
\begin{equation}\label{200}
(\Delta u)^2+(\Delta v)^2=\langle \hat a_1^\dagger,\hat a_1\rangle+\langle \hat a_1,\hat a_1^\dagger\rangle+\langle \hat a_2^\dagger,\hat a_2\rangle+\langle \hat a_2,\hat a_2^\dagger\rangle+2(\langle \hat a_1,\hat a_2\rangle+\langle \hat a_1^\dagger,\hat a_2^\dagger\rangle).
\end{equation}
In our system $\hat a_1=\hat a$ and $\hat a_2=b$, hence the above equation is rewritten as
\begin{equation}\label{201}
(\Delta u)^2+(\Delta v)^2=\langle \hat a^\dagger,\hat a\rangle+\langle \hat a,\hat a^\dagger\rangle+\langle \hat b^\dagger,\hat b\rangle+\langle \hat b,\hat b^\dagger\rangle+2(\langle \hat a,\hat b\rangle+\langle \hat a^\dagger,\hat b^\dagger\rangle),
\end{equation}
where,
\begin{equation}\label{202}
\langle \hat a,\hat b\rangle=\langle \hat a\hat b\rangle-\langle \hat a\rangle\langle\hat b\rangle.
\end{equation}
Inview of the above relation and using commutation relation
\begin{equation}\label{203}
[\hat a,\hat a^\dagger]=[\hat b,\hat b^\dagger]=2,
\end{equation}
Eq.~\eqref{201} takes the form
\begin{eqnarray}\label{204}
(\Delta u)^2+(\Delta v)^2&=&4+2\langle \hat a^\dagger\hat a\rangle-2\langle \hat a^\dagger\rangle\langle\hat a\rangle+2\langle \hat b^\dagger\hat b\rangle-2\langle \hat b^\dagger\rangle\langle\hat b\rangle\nonumber\\
&+&2(\langle \hat a\hat b\rangle-\langle \hat a\rangle\langle\hat b\rangle+\langle \hat a^\dagger\hat b^\dagger\rangle-\langle \hat a^\dagger\rangle\langle\hat b^\dagger\rangle).
\end{eqnarray}
With the help of Eqs.~\eqref{139} to ~\eqref{142} along with Eqs.~\eqref{58} to ~\eqref{65} and Eqs.~\eqref{51} and ~\eqref{52} we can show for
\begin{equation}\label{205}
\langle \hat a^\dagger\hat a\rangle=q^2+\langle\hat a'^\dagger\hat a'\rangle=q^2+\langle\hat a_2^\dagger\hat a_2\rangle.
\end{equation}
Similarly,
\begin{equation}\label{206}
\langle \hat b^\dagger\hat b\rangle=q^2+\langle\hat b_2^\dagger\hat b_2\rangle,
\end{equation}
\begin{equation}\label{207}
\langle \hat a^\dagger\rangle\langle\hat a\rangle=\langle \hat b\rangle\langle\hat b^\dagger\rangle=\langle \hat a\rangle\langle\hat b\rangle=\langle \hat a^\dagger\rangle\langle\hat b^\dagger\rangle=q^2,
\end{equation}
\begin{equation}\label{208}
\langle \hat a\hat b\rangle=q^2+\langle \hat a_2\hat b_2\rangle,
\end{equation}
\begin{equation}\label{209}
\langle \hat a^\dagger\hat b^\dagger\rangle=q^2+\langle \hat a_2^\dagger\hat b_2^\dagger\rangle.
\end{equation}
Hence using Eqs.~\eqref{205} to ~\eqref{209} into Eq.~\eqref{204} we observe that 
\begin{eqnarray}\label{210}
(\Delta u)^2+(\Delta v)^2=4+4\langle \hat a_2^\dagger\hat a_2\rangle+4\langle \hat a_2\hat b_2\rangle.
\end{eqnarray}
Substituting Eqs.~\eqref{157} and ~\eqref{162} in the above equation and rearranging yields
\begin{eqnarray}\label{211}
(\Delta u)^2+(\Delta v)^2&=&4-\frac{4\gamma}{\kappa+2\gamma}.
\end{eqnarray}
Compairing with Eq.~\eqref{175} we see that the sum of the fluctuation of the two EPR like operators equals the plus quadrature variance for a superposed two-mode coherent and subharmonic cavity lights. Thus,
\begin{eqnarray}\label{212}
(\Delta u)^2+(\Delta v)^2=(\Delta c_+)^2.
\end{eqnarray}
We finally see that at steady state and threshold ($\kappa=2\gamma$) the sum of the fluctuations of the two EPR like operators for a superposed two mode coherent and subharmonic cavity light turns out to be
\begin{eqnarray}\label{213}
(\Delta u)^2+(\Delta v)^2=3.
\end{eqnarray}
Therefore, inview of the above equation the inequality described in Eq.~\eqref{195} is satisfied. Which shows a superposed two-mode coherent and subharmonic cavity light beams are entangled at steady state and threshold.
\par
We now proceed to calculate degree of entanglement for a superposed two-mode light. For this purpose we define degree of entanglement as   
\begin{equation}\label{214}
D_E=\frac{(\Delta u)^2+(\Delta v)^2}{4}.
\end{equation}
Inview of Eq.~\eqref{213} we observe that a superposed two-mode coherent and subharmonic cavity light beams are $75\%$ entangled.
\begin{center}
\begin {figure}[h]
\centering
\includegraphics [height=10cm, width =14cm ]{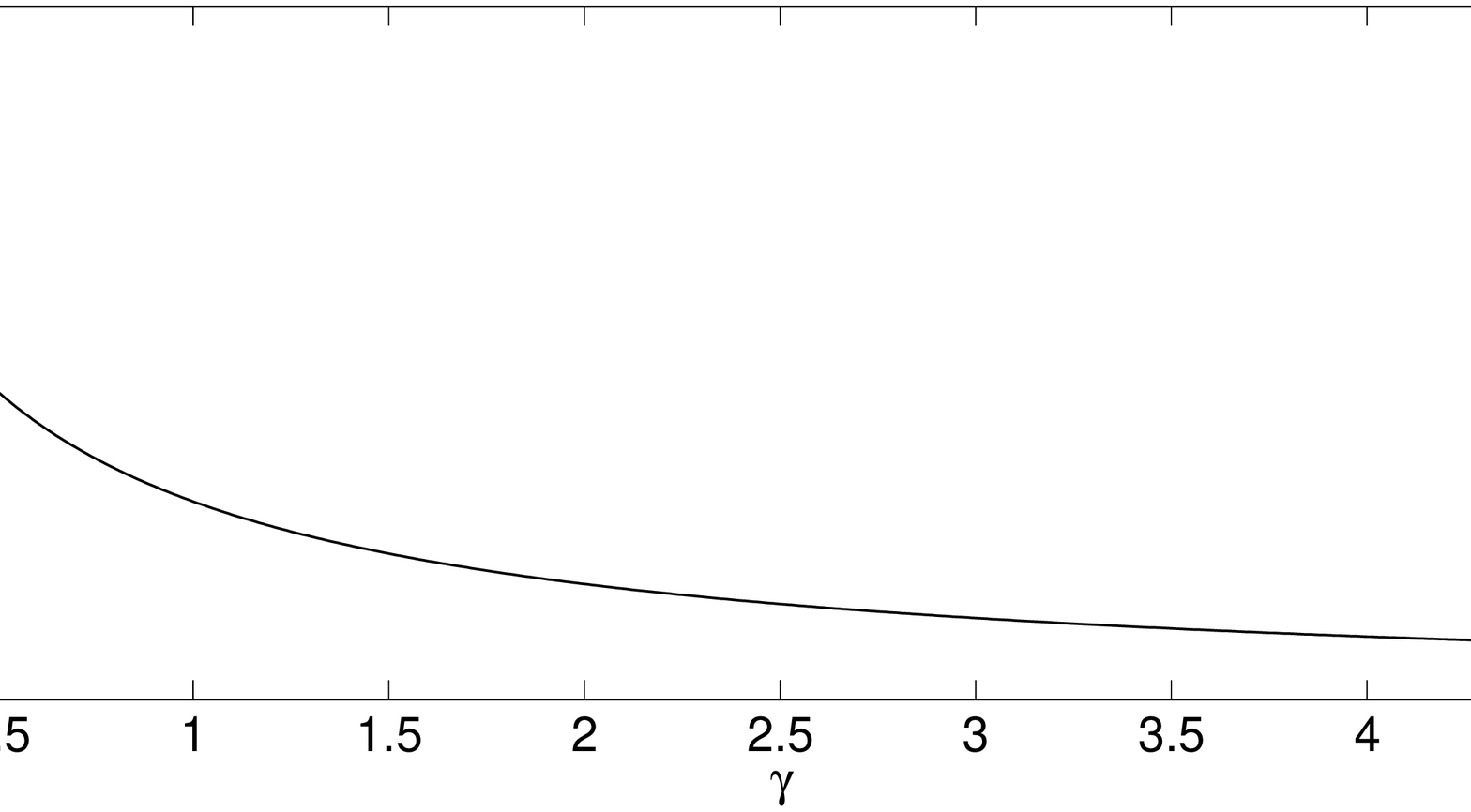}
\caption{ A plot of sum of fluctuation in EPR operators versus $\gamma$ for $\kappa=0.8$ }
\label {fig:p}
\end {figure}
\end {center}
\newpage
\section *{Conclusion}
We have analyzed Q function, photon statistics (mean photon number and variance of a photon number), quadrature statistics (quadrature variance and squeezing), and entanglement criterion (Cauchy-Schwarz and continuous variable(CV)).  In order to carry out our analysis, we tried to considered a quantum system with Gaussian operators with zero mean. We have obtained that a superposed two-mode coherent and subharmonic cavity light beams are in a squeezed state and the squeezing occurs in the plus quadrature. Besides, our analysis shows that at steady state and at threshold, a superposed two-mode coherent and subharmonic light beams have a maximum squeezing of $25\%$ below the two-mode vacuum-state level. We have also clearly shown that a superposed two-mode light beams are entangled at steady-state and threshold and the entanglement turned out to be observed in the highly correlated squeezed photons with $75\%$ degree of entanglement.

\end{document}